\documentclass[usenatbib]{mnras}
\pdfoutput=1
\usepackage[usenames,dvipsnames]{color}
\usepackage{float}  
\usepackage[pdftex]{graphicx}
\usepackage{xspace}
\usepackage{amssymb}
\usepackage{xcolor}
\makeatletter
\renewcommand{\@makefnmark}{\hbox{\textsuperscript{\scriptsize{\@thefnmark}}}}
\makeatother

\makeatletter
    \long\def\@makefntext#1{\parindent 1em\noindent
            \hb@xt@1.8em{%
                \hss\@textsuperscript{\scriptsize\@thefnmark}}#1}%
\makeatother

\newcommand{\scii}   {\,{\sc ii}}

\usepackage{url}

\usepackage{subcaption}

\def\arcmin{\hbox{$^\prime$}}
\def\kpc{\hbox{kpc}}

\def\arcsec{\hbox{$^{\prime\prime}$}\xspace}

\def\kms{km~s$^{-1}$}

\def\chan{{\it Chandra}\xspace}

\usepackage{titlesec}

\def\apjl{}

\def\aap{A\&A}

\newcommand{\Msun}{\hbox{$\hbox{M}_\odot\;$}}

\newcommand{\myemail}{mapt@iac.es}
\newcommand{\beq}{\begin{equation}}
\newcommand{\eeq}{\end{equation}}
\newcommand{\mr}{\mathrm}
\usepackage{amssymb,amsmath}
\usepackage{epsfig}
\usepackage{rotating}
\usepackage{times}
\usepackage{natbib,twoopt}

\definecolor{darkcyan}{rgb}{0.0, 0.55, 0.55}
\definecolor{brickred}{rgb}{0.8, 0.25, 0.33}

\usepackage{hyperref}
\hypersetup{colorlinks=True, linkcolor=darkcyan, urlcolor=brickred, citecolor=Orange}

\title[The nature of  CX1004]
  {Constraining the nature of the accreting binary in CXOGBS J174623.5-310550}
\author[Torres et al.] {
M. A. P. ~Torres$^{1,2,3}${\thanks{{\color{Purple}E-mail:\myemail}}}, 
S.~Repetto$^{4,5}$,  
T.~~Wevers$^{6,5,3}$, 
M.~~Heida$^{7}$, 
P.G.~~Jonker$^{3,5}$, 
R.I.~~Hynes$^8$,  \newauthor
G.~~Nelemans$^{5,9}$,  
Z. ~~Kostrzewa-Rutkowska$^{3,5}$,
L. ~~Wyrzykowski$^{10}$,
C.T.~~Britt$^{11}$, \newauthor
C.O.~~Heinke$^{12,13}$, 
J. ~~ Casares$^{1,2}$,
C.B.~~Johnson$^{8}$, 
T.J.~~Maccarone$^{11}$, 
D.T.H.~~Steeghs$^{14}$
  \\
$^{1}$Instituto de Astrof\'{i}sica de Canarias, V\'{i}a L\'{a}ctea, La Laguna, E-38205, Santa Cruz de Tenerife, Spain\\
$^{2}$Departamento de Astrof\'{i}sica, Universidad de La Laguna, E-38206, Santa Cruz de Tenerife, Spain\\
 $^3$ SRON, Netherlands Institute for Space Research, Sorbonnelaan 2, 3584 CA, Utrecht, The Netherlands \\
$^4$Physics Department, Technion - Israel Institute of Technology, Haifa, Israel 32000\\
$^5$Department of Astrophysics/IMAPP, Radboud University, P.O. Box 9010, 6500 GL Nijmegen, The Netherlands.\\
   $^6$ Institute of Astronomy, Madingley Road, Cambridge CB3 0HA, United Kingdom\\
    $^7$ Space Radiation Laboratory, California Institute of
    Technology, Pasadena, CA 91125, USA\\
     $^8$ Department of Physics and Astronomy, Louisiana State University, Baton Rouge, LA 70803, USA\\
    $^9$ Institute for Astronomy, KU Leuven, Celestijnenlaan 200D, 3001 Leuven, Belgium\\
        $^{10}$ Warsaw University Astronomical Observatory, Al. Ujazdowskie 4, 00-478 Warszawa, Poland\\
    $^{11}$ Department of Physics \& Astronomy, Texas Tech University, Box 41051, Lubbock ,TX 79409-1051, USA\\
    $^{12}$ Department of Physics, University of Alberta, CCIS 4-183, Edmonton, AB T6G 2E1, Canada\\
    $^{13}$ Max Planck Institute for Radio Astronomy, Auf dem Hugel 69, D-53121 Bonn, Germany\\
    $^{14}$ Department of Physics, University of Warwick, Coventry CV4
    7AL, UK }

\date{Accepted XXX. Received XXX}
\pagerange{\pageref{firstpage}--\pageref{lastpage}} \pubyear{0000}
\pagerange{\pageref{firstpage}--\pageref{lastpage}} \pubyear{2014}
\def\LaTeX{L\kern-.36em\raise.3ex\hbox{a}\kern-.15em
    T\kern-.1667em\lower.7ex\hbox{E}\kern-.125emX}

\begin{document}

\label{firstpage}

\maketitle
\label{firstpage}
\begin{abstract}
We report optical and infrared observations of the X-ray
source CXOGBS J174623.5-310550. This Galactic object was identified as  a
potential quiescent low-mass X-ray binary accreting from an M-type donor
on the basis of optical spectroscopy and the broad H$\alpha$ emission
line.  The analysis of X-shooter spectroscopy covering 3
consecutive nights supports an M2/3-type spectral classification. Neither radial velocity variations nor rotational
broadening is detected in the photospheric lines.  No periodic variability 
is found in I- and $r^\prime$-band light curves. We derive
$r^\prime = 20.8$, I = 19.2 and $K_\mr{s}\approx16.6$ for the optical and
infrared counterparts with the  M-type star contributing $\approx90\%$ to
the I-band light. We estimate its distance to be $1.3-1.8$ \kpc. The lack of
radial velocity variations implies that the M-type star
is not the donor star in the X-ray binary. This could be an
interloper or the outer body in a hierarchical triple. We constrain
the  accreting binary  to be a $\lesssim 2.2$ hr
 orbital period eclipsing  cataclysmic variable or a low-mass X-ray binary
 lying in the foreground of the Galactic Bulge.

\end{abstract}


\section{Introduction}

Interacting binaries containing compact objects provide a means to study
the evolution of stars in binaries, and in particular the formation
of their most compact remnants: neutron stars or black holes (BHs).
The current Galactic population of accreting stellar-mass BHs amounts to $18$
objects with a dynamical mass measurement
(\citealt{2014SSRv..183..223C};
\citealt{2016A&A...587A..61C}). Increasing the sample of accreting Galactic BHs is of great importance for three main reasons. Firstly, different supernova models predict a
different BH mass distribution (\citealt{2001ApJ...554..548F}; \citealt{2012ApJ...757...91B}; \citealt{2012ApJ...749...91F}; \citealt{2012ApJ...757...69U}). Therefore, an unbiased determination of the BH mass distribution can be used to constrain supernova models.
Secondly, increasing the sample of BHs with measured space velocities would help
in the long debate on the type of natal kicks received by BHs at
formation, which again provides input to supernova and binary evolution models
(\citealt{2004MNRAS.354..355J}; \citealt{2014PASA...31...16M};
\citealt{2015MNRAS.453.3341R}; \citealt{mine}). Thirdly, the comparison between the
observed number of accreting BH binaries and the predicted number from population
synthesis models, would help in unraveling the physics involved in the formation and
evolution of these sources (\citealt{1992ApJ...399..621R}; \citealt{1997A&A...321..207P}; \citealt{1998ApJ...493..351K}; \citealt{2003ApJ...597.1036P};  \citealt{2004ApJ...603..690B}; \citealt{2006MNRAS.369.1152K}; \citealt{2006A&A...454..559Y}). 

In an attempt to enlarge the sample of compact X-ray  binaries and thus
address the above questions, the {\it Chandra} Galactic Bulge Survey
(GBS) imaged in X-rays a 12 deg$^2$ area towards the Bulge with a limiting
sensitivity set to maximize the number of detected quiescent
low-mass X-ray binaries (LMXBs) over cataclysmic variables (CVs; see
\citealt{2011ApJS..194...18J, 2014ApJS..210...18J} for complete
details on the GBS design). The X-ray survey has been complemented
with dedicated multi-frequency and variability studies, e.g.
\citet{2012MNRAS.426.3057M}, 
\citet{2012ApJ...761..162H},
\citet{2012AcA....62..133U},
\citet{2014MNRAS.438.2839G},
\citet{2014ApJS..214...10B},
\citet{2016MNRAS.458.4530W},
\citet{2017MNRAS.466..163W}. Such studies are facilitating the
identification of multiband counterparts to the
1640 X-ray sources found in the GBS. These counterparts are being
classified on the basis of their spectroscopic and photometric properties (see 
\citealt{2012MNRAS.426.3057M},  
\citealt{2013ApJ...769..120B}, 
\citealt{2014MNRAS.440..365T}). 
Further photometry and spectroscopy is performed for  candidate accreting binaries
in order to establish the nature of their accretors (e.g.
\citealt{2013MNRAS.428.3543R},
\citealt{2016MNRAS.462L.106W},
\citealt{2017MNRAS.466..129J}). In this paper we present the follow-up observations of CXOGBS
J174623.5-310550 (referred to hereafter as CX1004). CX1004 was detected
with three $0.3-8$ keV  counts  during the \chan GBS survey \citep{2011ApJS..194...18J}, implying a  $0.5-10$ keV X-ray luminosity of $L_x (d) \sim 2 \times 10^{31} \times
(\frac{d}{1.0~\mr{kpc}})^2$ erg s$^{-1}$. Its optical counterpart
was found at $r^\prime = 20.75 \pm 0.02$ and $i^\prime = 19.35
\pm 0.01$ in observations taken with  the 4-m
Victor M. Blanco telescope camera Mosaic-II in June 2006 \citep{2016MNRAS.458.4530W}. A subsequent Mosaic-II $r^\prime$-band light
curve obtained during  July 2010 did not show any significant photometric variability on a time-span of $8$ days
(\citealt{2014MNRAS.440..365T}, \citealt{2014ApJS..214...10B}).
Optical spectroscopy taken with the VIsible Multi-Object Spectrograph
(VIMOS) at the Very Large Telescope (VLT) and with the Gemini
Multi-Object Spectrograph (GMOS) were presented in \citet{2014MNRAS.440..365T} and \citet{2015MNRAS.448.1900W} .
The data showed absorption features  consistent with those of an early M-type star
and a broad double-peaked $\mr{H}\alpha$ emission line with $2120 -
2500$ \kms full-width half-maximum (FWHM),~flagging CX1004 as an
accreting binary, either a low accretion rate high-inclination
CV  or a quiescent LMXB.

This paper is organized as follows: the
  observations and data reduction steps are detailed 
 in sections \ref{sec:xsh-data} and \ref{photometric_data}, respectively. In section
 \ref{sec:analysis} the optical spectra and the optical/infrared photometry
 are analyzed. In section \ref{sec:discussion} a discussion of the
 results is presented. Our conclusions are drawn in section \ref{sec:conclusions}.

\section{Spectroscopic data}
\label{sec:xsh-data}
Time-resolved spectroscopy of CX1004 was obtained using the medium resolution X-shooter echelle spectrograph
(\citealt{2011A&A...536A.105V}) mounted at the 8.2-m ESO Unit 2
VLT. The observations were obtained under program 088.D-0096(A). X-shooter provides spectra covering a large wavelength
range of $3000-24800$ \AA, split into three spectroscopic  arms: UVB, VIS and NIR.  For our analysis we focus on the VIS and UVB data, which cover the range $\lambda\lambda3000-10240$ with a dispersion of $\sim0.2$ \AA~pixel$^{-1}$ in both arms. The NIR-arm
data were not used due to their lower signal-to-noise ratio (SNR) and the light contamination from a nearby field star. The observations were taken with a slit width
of $0\arcsec.9$ in the VIS arm and of $1\arcsec.0$ in the UVB arm
which delivered a resolving power of $\simeq 8800$ and $\simeq 5100$, respectively. We executed observing blocks consisting of an ABBA nodding sequence, with integration times
for each spectrum of $900$ s (VIS) and $877$ s (UVB). 
To reduce
systematic effects due to possible excursions of the target position
with respect to the slit center, we re-acquired CX1004 at the start of
each one hour-long observing block. We also observed with the same setup and in nodding mode three red
dwarfs that we will use for the analysis presented in section
\ref{sec:abs}: GJ 9592,  GJ 465 and GJ 402 with
spectral types M1, M2 and M4, respectively.\\
\indent Six, eight and six spectra were collected on 1 March 2012 from airmass $1.6$ to $1.2$, on 2 March 2012 from airmass $1.9$ to $1.2$ and on 3 March 2012  
from airmass $1.6$ to $1.2$, respectively. From the FWHM of the
collapsed spatial profile of the source spectrum at spectral positions close to $\lambda6300$ we measure
a mean image quality ranging from $1\arcsec.7$ to $1\arcsec.4$,
$3\arcsec.4$ to $2\arcsec.0$ and $3\arcsec.2$ to $1\arcsec.5$ for the
first, second and third night, respectively. A similar FWHM was found
at positions near $\lambda9000$. Therefore, the VIS and UVB data were
obtained in slit-limited conditions. On the other hand,  because of  the faintness of the
source and the poor seeing during the observations,
the individual spectra have low SNR: $\sim2-5$ near
H$\alpha$ and $\sim10$ in regions covering the Ca triplet.  \\

We reduced each individual $900$ s VIS and each $877$ s UVB
frame in order to optimize the time resolution. We processed the data
using  {{\small{{\texttt{EsoRex}}}}, a software
  package delivered within the X-shooter pipeline
  (\citealt{2010SPIE.7737E..28M}). In this way, the data were bias and flat
  field corrected, the echelle orders were merged and rectified, and
  the spectra wavelength calibrated. After several tests to investigate how to maximize the signal-to-noise of the extracted 1D spectra,  we performed the
  extraction of each 2D spectrum with {\small{\texttt{IRAF}}}} using an
  extraction aperture with size equal to the FWHM measured from the
  spatial profile of the spectrum in question. The resulting extracted
  spectra of CX1004 and the spectral-type templates were subsequently imported in {\small{{\texttt{MOLLY}}}}, rebinned to a uniform pixel scale. We checked the zero point of the wavelength calibration of our spectra
measuring the velocity shift  of the 
sky emission lines [OI] $\lambda\lambda 6300.304, 8310.719$
(\citealt{1996PASP..108..277O}). We used these shifts  to correct for
the zero-point deviations in the spectral regions covering H$\alpha$ and the Ca{\scii} infrared triplet. 
The median offsets in the $\lambda6300$ and $\lambda8311$ sky  lines
were $\leq3.6$ and $\leq4.3$ \kms\, in amplitude, respectively. No
suitable sky emission lines were available in the spectral range
covered by the UVB arm, thereby a zero-point correction was not
possible. Finally, the spectra were corrected for the motion of the
Earth.

\section{Photometric data}
\label{photometric_data}

\subsection{GBS optical point source catalogue}
\label{sec:GBScat}

CX1004 is found in the GBS optical point source catalogue
\citep{2016MNRAS.458.4530W} which  consists of  optical photometry
obtained using the Mosaic-II imager on the 4-m  Victor M. Blanco
telescope at CTIO.  With 8 CCDs, the instrument covered a $36\arcmin\times
36\arcmin$ field of view (FOV) with a  plate scale of
$0\arcsec.27$~pixel$^{-1}$. The photometry was taken
between 21 and 29
   June 2016.  The GBS area was covered in the
   $r^\prime$, $i^\prime$ and $\mr{H}\alpha$ filters, with exposure
   times of  $120$, $180$ and $480$ s respectively.  The data
   reduction steps, as well as the photometric and astrometric calibration, are described in detail in \citet{2016MNRAS.458.4530W}. This catalogue
is our  photometric and astrometic reference or the
  calibration of the new data
presented in this paper  (sections \ref{sec:decamdata},
\ref{sec:fors2data}) and for the recalibration of other, previously published photometry (section \ref{sec:OGLElc}).

\subsection{OGLE}
CX1004 is listed as object ID 18319 in the Bulge field BLG659.29  monitored  during the
fourth phase of the  Optical Gravitational Lensing Experiment (OGLE-IV).
Between 23 April 2010 and 19 March 2011
a total of $60$ I-band data points were collected with the 1.3-m
Warsaw telescope located at Las Campanas observatory under a $\approx
1\arcsec.3$ seeing sampled with a plate
scale of $0\arcsec.26$ pixel$^{-1}$. Photometry was obtained using the
difference imaging technique
tied to the OGLE data (see \citealt{2015AcA....65....1U} for a detailed overview
of the OGLE-IV survey). 

\subsection{VVV}
Archival infrared images from the VISTA  Variables in the Via Lactea
Survey (VVV, \citealt{2010NewA...15..433M}) were inspected to search
for the infrared counterpart to CX1004. The survey data were obtained 
with the 4-m VISTA telescope at
Paranal Observatory that made use of VIRCAM, a camera with a $1.1 \times
1.5$ deg$^2$ FOV and a plate scale of $0\arcsec.34$ pixel$^{-1}$. In our
study we used data acquired in the
$K_s$-band during 15 August 2012 and 23 March 2013 with total time
on-source of 48 s and an image quality better than $0\arcsec.8$. 

\subsection{DECam}
\label{sec:decamdata}
As a continuation of the GBS variability survey published in \citet{2014ApJS..214...10B},
time-resolved photometry of the field containing CX1004 was obtained
on 10-11 June 2013 with the Dark Energy CAMera  (DECam;
\citealt{2008SPIE.7014E..0ED}) on the 4-m Victor M. Blanco telescope
at CTIO. DECam uses a mosaic of 62 CDD each with 2k $\times$ 4k pixels to cover a 2.2
deg$^2$ FOV with a   plate scale of $0\arcsec.27$ pixel$^{-1}$. A
total of $109$ $r^\prime$-band images were taken with an exposure time
of $90$ s each. The seeing was between $0\arcsec.8-2\arcsec.5$ over both nights.
The images were reduced with the NOAO DECam pipeline and astrometry
was performed using the pipeline WCS on the re-projected images. Instrumental magnitudes were extracted using point spread
function (PSF)  photometry with the  {\small\texttt{DAOPHOT}} task in
{\small{\texttt{IRAF}}}.  Finally, the  photometric calibration was
performed against the GBS point source catalogue (section \ref{sec:GBScat}).  

\subsection{FORS2}
\label{sec:fors2data}
Single B- and I-band images of the field containing CX1004 were
obtained with the FOcal Reducer and low dispersion Spectrograph 2
(FORS2; \citealt{1998Msngr..94....1A}) mounted on the 8.2-m ESO Unit 1 VLT.  The observations were obtained under program 095.D-0973(A) during 12 June 2015.  The
instrument was used with the standard resolution collimator
($6\arcmin.8 \times 6\arcmin.8$ FOV) and the
mosaic of two $2 \times 4$ k MIT CCDs. The CCDs were binned $2
\times 2$ providing a plate scale of $0\arcsec.25$ pixel$^{-1}$.  The
integration times were $300$ s and $30$ s for the images taken with
the B- and I-Bessel filters, respectively. The image quality was
$0\arcsec.9$ (B-band) and $0\arcsec.75$ FWHM (I-band). The data were
bias subtracted and flat-field corrected using standard tasks in
{\small{\texttt{IRAF}}}.  The computed astrometric solution for
the images had an r.m.s of  $\approx 0\arcsec.162$. Then we used
{\small{{{\texttt{IRAF}}}}} tools to shift the images to the same reference frame using the
centroids of 103 point sources in the images. 
   
   \begin{table*}
\caption{{Photometric and spectroscopic data. The Mosaic-II, VIMOS and
    GMOS data sets reanalyzed
    in this work are published first in \citet{2014ApJS..214...10B}, 
\citet{2014MNRAS.440..365T} and \citet{2015MNRAS.448.1900W}, respectively.}}
\label{tab:Journal}
    \begin{tabular}{llcccccccc} 
        \hline Instrument & Date & \# & Exp. time  &  Spec. range 
& Seeing & Slit
        width & Resol. & Disp. & Plate scale\\ 
       /survey &  &  & (s) &(\AA/band)
& (\arcsec) & (\arcsec) & (\AA) &
        (\AA~pix$^{-1}$) & (\arcsec~pix$^{-1}$)\\ 
        \hline
        OGLE & 2010/04/23  & 60 & 100 & I & 1.3 &  - & - & - & 0.26 \\
        &-2011/03/19 & &&&&&& &\\
         MOSAIC-II & 2010/07/12-18 &37 & 120 & $r^\prime$ & 1.0 & - &
                                                                      -
                       & - & 0.26 \\
          VIMOS &  2011/06/28  & 2 &875 & 4800-10000 & 1.2 & 1.0 & 10 & 2.5& 0.2  \\
        X-shooter & 2012/03/1,2,3  & 6,8,6 & $900$ &5595-10240 & 1.6, 2.7, 2.4  & 0.9 & $\sim1 $ & $\sim0.2 $  & $\sim0.3$ \\
 &  & 6,8,6 & $877$ &3000-5595 &  -  & 1.0 & $\sim1$ & $\sim0.2$ &  $\sim0.3$ \\
      GMOS  &   2012/05/14,17 & 12 & $900$ &4800-7600 & 0.8, 0.75 & 0.75 &   5 & 1.36 & 0.15  \\
      VVV & 2012/08/15, & 1 & $48$ & $K_s$ & $0.7$ &  - & - & - & $0.34$ \\
      & 2013/03/22         & 1 &        &  & $0.8$&&&&\\
      DECam &  2013/06/10,11  & 109 & $90$ &$r^\prime$  & $0.8-2.5$ &  - & - & - & $0.27$ \\
                FORS2 & 2015/06/12  & 1,1 & 300, 30 & B,I& 0.9, 0.75 & - &  -& -   & $0.25$ \\ 
                   \hline
    \end{tabular}
    \newline
    \end{table*}

\section{Data Analysis and Results}
\label{sec:analysis}
\subsection{Determining the radial velocities, spectral type and rotational broadening of the M-type star}
\label{sec:abs}
We measure the radial velocities of the candidate counterpart,
cross-correlating its X-shooter spectra with those of a template star \citep{1979AJ.....84.1511T}. Prior to the cross-correlation, target and template
spectra were  resampled into a common logarithmic wavelength
scale. Next, all spectra were normalized over the range $\lambda\lambda8010-8810$
by dividing them with the result of a third-order
spline function fit to
the continuum obtained while masking strong spectral features. The normalized wavelength range contains the resolved Ca{\scii}
infrared triplet that, in contrast to other photospheric
lines, is detected in most of the individual spectra  despite their low
SNR. The Na I doublet is also evident, but we choose not to include it in the analysis since it is contaminated by telluric lines.
 All computed cross-correlation functions showed
 a significant peak, with best results achieved when using only the
 wavelength intervals covering the sharp Ca{\scii} triplet lines. The
radial velocity values provided in this paper were obtained by 
cross-correlating the target data against the M2-type template, which
best matches the spectral type for CX1004 (see
  below). We corrected the resulting velocities for the
intrinsic $51.17$ \kms \, systemic radial 
velocity of the M2 dwarf, which is  accurate to approximately 0.1 \kms\,
(\citealt{2002ApJS..141..503N}). The resulting radial velocities do
not show any significant variations during the three nights of
observations. Their nightly means and
standard deviations are: $-23\pm 3$, $-25  \pm  5$~and $-24 \pm
3$ \kms~ for 1, 2 and 3 of March, respectively. 

%\begin{figure*}
%\centering \includegraphics[width=4.5in,angle=90]{fig1.pdf}
%\caption{Radial velocities derived by
 % cross-correlating the 20 CX1004 spectra in the Ca{\scii} triplet region with the M2 V t % emplate.}
%\label{fig:XCOR}
%\end{figure*}

\begin{figure}
\centering \includegraphics[width=1.\columnwidth]{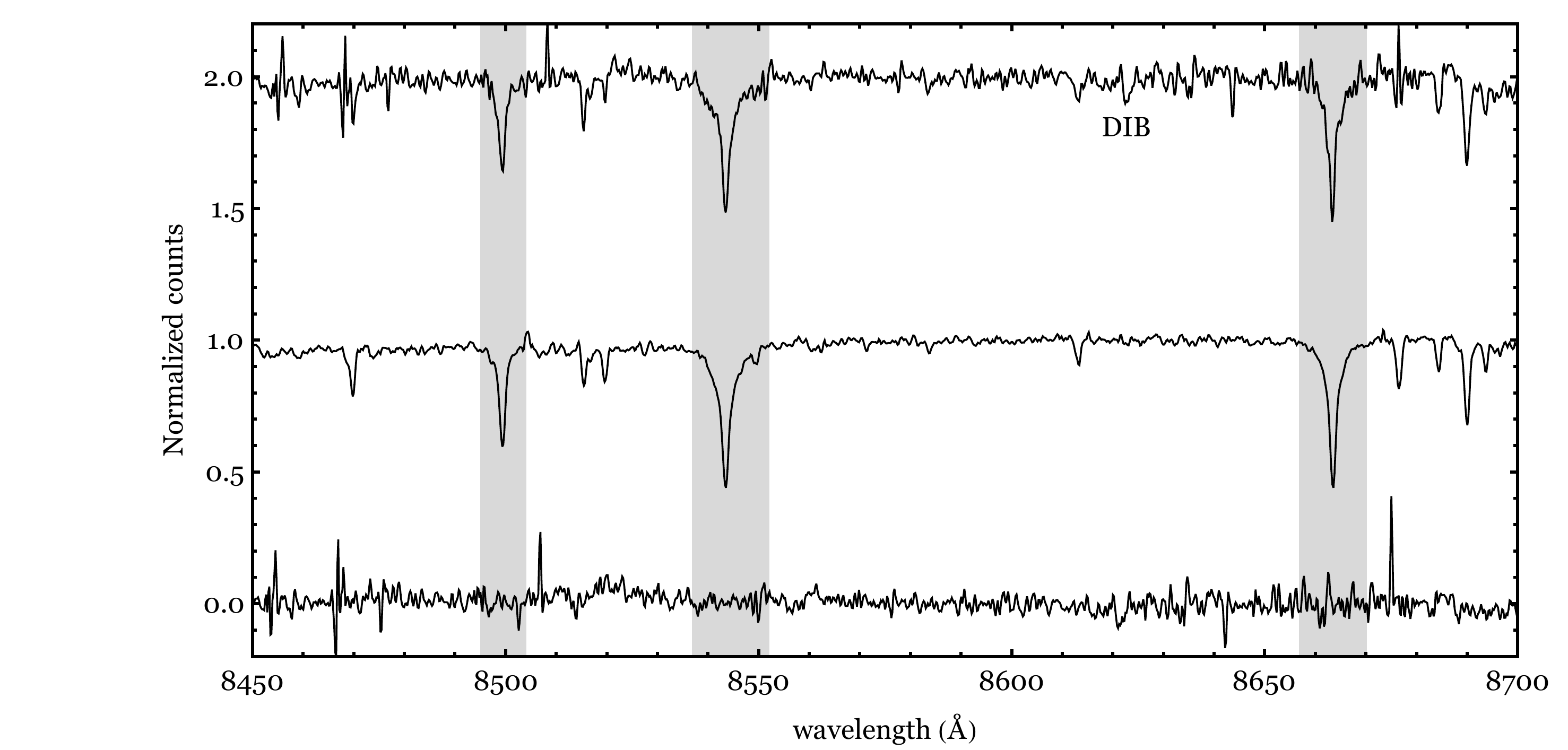}
\caption{From top to bottom: the normalized velocity--corrected average
  spectrum of CX1004, the M2 template, and the residuals after optimal subtraction. Arbitrary vertical offsets have been applied to the spectra for the sake of clarity.
The shaded areas mark the Ca{\scii} triplet absorption lines. The $\lambda8621$  DIB is also indicated.}
\label{fig:Red}
\end{figure}

%\subsection{Spectral type and rotational broadening}
%\label{sec:Opt}
\citet{2014MNRAS.440..365T} supported an early M-type classification
for the optical counterpart to CX1004 on the basis of the presence of
prominent TiO $\alpha,\beta,\gamma$, $\gamma^{'}$ band systems and the
lack of the TiO $\delta$, $\epsilon$ band systems redward of $8000$
\AA\, in VIMOS spectra. The X-shooter spectra confirm these
results. In particular, there is no evidence for TiO molecular bands
in the range $\lambda\lambda8200-8800$ (see Fig. \ref{fig:Red}).
Molecular bands start to be evident in this region for spectral types
later than M3 (see e.g. \citealt{1984ApJ...283..457J};
\citealt{2015AJ....150...42Z}).To verify our visual classification,
that does not account for the possible contribution from an accretion
flow to the optical continuum, we apply the optimal subtraction method
described in \citet{1994MNRAS.266..137M}. This method allows to
determine the spectral type of the stellar component, its fractional
contribution to the total light and the rotational broadening of the
photospheric lines.  We focused this analysis on the Ca{\scii}
infrared triplet region where
the SNR is higher.

First, the normalized target spectra were velocity-shifted to the rest
frame of the template star by subtracting the radial velocities
obtained from the cross-correlation with the template. Next, the
velocity-shifted spectra were averaged, with different weights to
maximize the SNR of the resulting sum.  The spectral templates were
then broadened from $0$ \kms~ to $100$ \kms~ in steps of $2$ \kms
through convolution with the rotational profile of
\citet{1992oasp.book.....G} adopting a limb darkening coefficient of
$0.75$.  Each broadened version of the template spectrum was
multiplied by a varying factor $f$ (representing the fractional
contribution of light from the template star) and next subtracted from
the CX1004 average spectrum. Then a $\chi^2$ test on the residuals was
performed to find the optimal value of $f $ by minimizing the $\chi^2$
between the residual of the subtraction and a smoothed version of
itself. We took the average of the $\chi^2$ and $f$ values obtained by
smoothing the residual using a Gaussian with FWHM from $25$ to $100$
\AA.  For each template, we are then able to produce a $\chi^2$-curve
as a function of the applied broadening, whose minimum provides
$V\sin{i}$ and $f$. The $V\sin{i}$ errors were obtained following the
bootstrapping approach outlined in \citet{2007ApJ...669L..85S}.  We
find that only in the case of the M4 V template the $\chi^2$-curve is
non-monotonic with a single minimum, whereas for the M1 and M2 dwarf
template, the $\chi^2$-curve is increasing monotonically with
$V\sin{i}$.  From the $\chi^2$-values (see Table \ref{tab:OptSub}),
we derive that the M2 dwarf provides the best fit to the averaged spectrum, while the
M4 template yields the highest $\chi^2$ and it has a non-physical value for
$f$. Therefore, we conclude the M2 template best matches the spectrum
of CX1004, although we cannot exclude a M3 spectral type for the star.  In
Fig. \ref{fig:Red} we show the average CX1004 spectrum (top), the M2
template (middle) and the residual after the optimal subtraction
(bottom). Both the averaged and residual spectrum show a diffuse
interstellar band (DIB) at $\lambda8621$. We will use its equivalent
width (EW) of $0.19 \pm 0.06$ \AA~to estimate the reddening in section
\ref{sec:Reddening}. Our analysis shows that the M2 template
contributes $\approx90\%$ to the total I-band flux. The results also
imply a low rotational broadening for the photospheric lines
$V\sin{i}\lesssim 35$ \kms,~an upper limit set by the spectral
resolution measured in the Ca {\sc ii} triplet
region.  In the remainder of this paper we will refer to the M2/3 star
associated to the optical counterpart as the M2 star or M2 companion.

\begin{table}
\caption{Spectral classification and rotational broadening. $\mr{d.o.f.} =$ degrees of freedom.}
\label{tab:OptSub}
\begin{center}
    \begin{tabular}{*{5}{| c}|} 
        \hline Template & Spectral & $V\sin{i}$  & $f$ & $\chi^2_{min}/\mr{d.o.f}$ \\

        &Type & \small{(\kms)} &
     &($\mr{d.o.f}\approx 160$)  \\
     \hline
      GJ 9592 & M1 V & - & $1.00\pm 0.05$ & $8.4$ \\
      GJ 465 &  M2 V & - & $0.90\pm0.01$ & $7.8$\\
     GJ 402 & M4 V & $19\pm 4$ & $1.20\pm 0.08$& $9.6$ \\

 \hline
    \end{tabular}
    \newline
    \end{center}
    \end{table}

\subsection{The properties of detected emission lines}
\label{sec:emission}

\begin{figure}
\centering \includegraphics[width=0.7\columnwidth]{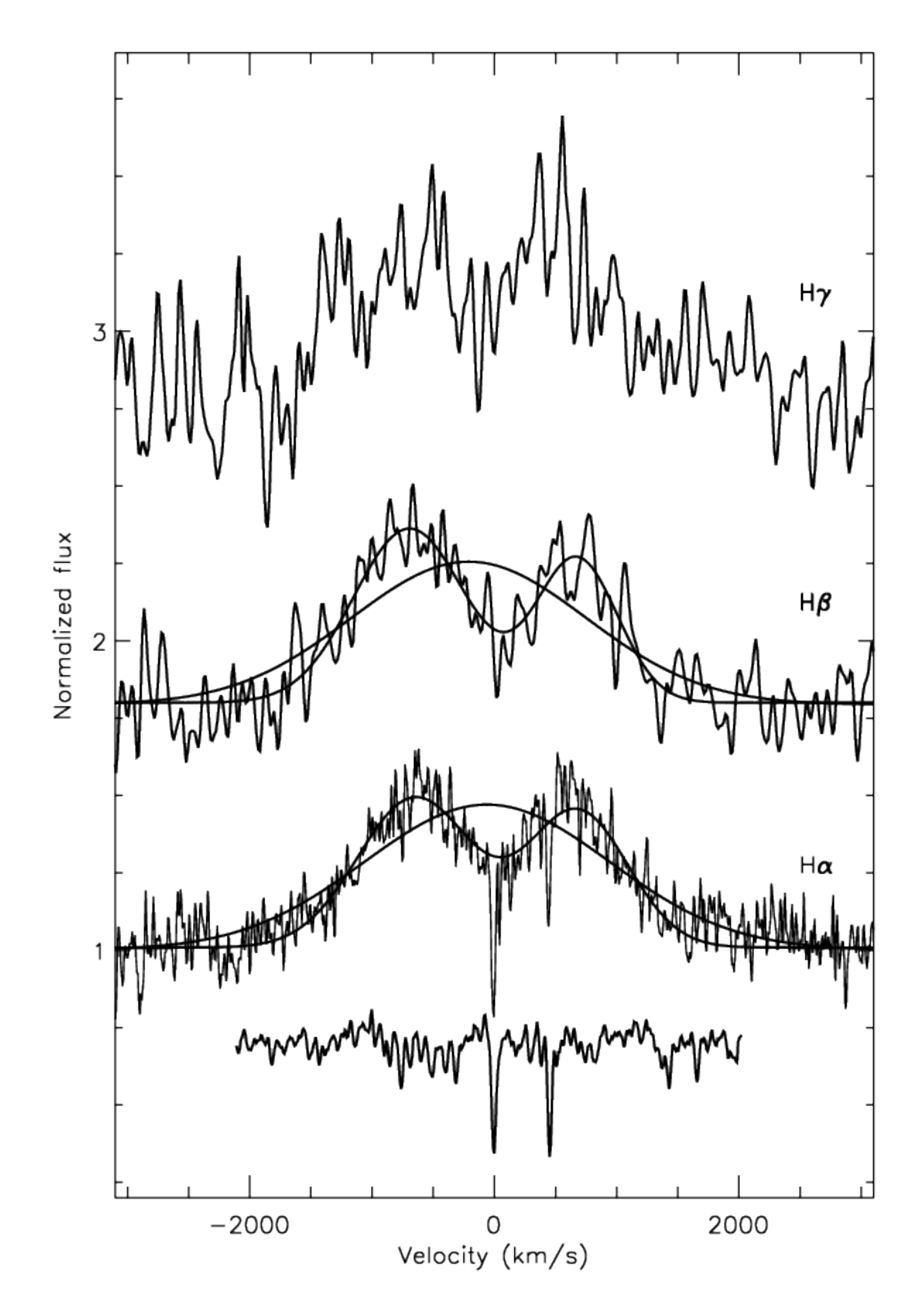}
\caption{Average profile of the Balmer emission lines in the
  X--shooter spectra of CX1004.  The spectra of the H$\beta$ and H$\gamma$
lines are smoothed with a 3--pixel--wide (FWHM) Gaussian. The
single and double-Gaussian fits performed in section \ref{sec:emission} are
overplotted. The bottom spectrum shows the M2 template where the strongest
photospheric absorption lines correspond to  H$\alpha$ and Ca \textsc{i} $\lambda6573$.}
\label{fig:Halpha}
\end{figure}

H$\alpha$  is the only emission line present in the X-shooter VIS
part of the spectra.  Its double-peaked morphology is apparent 
in the individual spectra with the highest SNR. H$\beta$ and H$\gamma$
emission lines are only detected after averaging the UVB
data (see Fig. \ref{fig:Halpha}). To characterize the H$\alpha$
emission line profile,  we first 
normalized the spectra by fitting the continuum adjacent to the line
with a low-order spline function. Next,
we produced nightly average spectra to increase the SNR. In this
process, we discarded data being consistent with noise in the
wavelength interval of interest - five and one spectra obtained in the
second  and third night, respectively. The resulting spectrum is shown in
Fig. \ref{fig:Halpha} together with the normalized
M2 template spectrum. By comparing them, it becomes evident that the
two more prominent narrow absorption features in the double-peaked
H$\alpha$ line are also present in the stellar template. We identify
them as  photospheric H$\alpha$ and Ca\textsc{i} $\lambda6573$ that
originate in the M2 star. Thus, the  X-shooter
  resolving power allows us to exclude the  possibility that the deep
  narrow H$\alpha$ absorption  is due to an inclination effect on the shape of the emission  profile. Such effect is frequently seen in eclipsing CVs and also in high
inclination BH LMXBs (e.g. \citealt{1987MNRAS.225..551M},
\citealt{2015MNRAS.450.4292T}). In what follows, we
will mask this absorption feature when fitting the average double-peaked profiles. 

Single and 2-Gaussian profiles were fit to the emission line to measure
its FWHM and the velocity shift of the blue ($V_b$) and red ($V_r$) peaks with
respect to the line's rest wavelength. The $(V_r+V_b)/2$ value provides
the centroid radial velocity (RV) with respect to the line rest wavelength
while $|V_r-V_b|$ yields the peak-to-peak separation, $\Delta V^{pp}$. 
The results from the fits are given in Table~\ref{tab:tabFWHM} together
with the EW of the lines. The uncertainties in the EWs
  were estimated from the scatter in the values obtained by using
  different wavelength intervals to place the local continuum level.
We deem the difference between the RV values obtained for the
per--night average X-shooter spectra to be due to noise on the line
structure rather than to RV changes in
the line centroid. The first night of observations has in general the
highest SNR spectra thus dominating the resulting weighted  average of
the 14 line profiles. Note that fits performed
without masking the absorption component from the M2 star,  yield
larger FWHMs (by $\sim100$ \kms) while the $\Delta V^{pp}$ and DP
values increase by $\lesssim2\%$. The EW measured for
  H$\alpha$ are low compared to those observed in quiescent CVs and
LMXBs. This is solely due to the strong contribution of the M2 star
to the optical continuum.  While the per--night average EW from the
X-shooter data are the same within the errors, they
are $30\%$ lower than found from the GMOS observations indicating
long-term variability of the line strength. We also
normalized the wavelength interval containing H$\beta$ and performed
single and 2-Gaussian model fits for the average profile resulting
from combining the three nights of data. We provide the
results from these fits in Table \ref{tab:tabFWHM}. The H$\beta$  EW measurement
is very uncertain given that the line continuum is difficult to establish. Finally,  $\mr{H}\gamma$ is marginally detected in the
averaged data and reliable fits were not possible.

 \begin{table*}
\caption{Best--fit parameters for the emission lines. The FWHMs are derived from a
  single Gaussian fit; while the line centroid radial velocity (RV)
  and peak-to-peak separation ($\Delta V^{pp}$) are obtained  from  a
  2-Gaussian fit to the emission line. DP is the double-peak
  separation measured from the 2-Gaussian fit when both components
are forced to have the same height and FWHM.}
\label{tab:tabFWHM}
    \begin{tabular}{lcccccc} 
 \hline Line & Night & \#& FWHM & RV & $\Delta V^{pp}$,(DP) & EW   \\ 
   Instrument  & Average & spectra & (\kms) & (\kms)
                                   &(\kms) & (\AA)\\ \hline
       \multicolumn{7}{|l|}{$\mathrm{H}\alpha$ $\lambda6562.760$}  \\ 
     &&&&& \\
        X-shooter & 1  March 2012  &6&$2330 \pm 20$ & $-10 \pm 5$ & $
         1310 \pm 10$& $-22 \pm 2$ \\ 
         & 2 March 2012  &3& $2310 \pm 40$ & $-90 \pm 10$ & $1370 \pm  20$ &$-25   \pm 3$\\ 
         & 3 March 2012  &5&$2460 \pm 30$ & $30 \pm 10$ & $1340 \pm 20$
                                                         & $-22 \pm 3$ \\
         & 1,2,3 March 2012  &14&$ 2360 \pm 20$ &$-10 \pm 5$ & $1330 \pm10$, &$-22 \pm 2$  \\ 
         &&   &  &  & $(1320 \pm 10$) & \\ 

        GMOS & 14, 17  May 2012  &12& $2350 \pm 40$ & $ -20 \pm 20$&
                                                                  $1350 \pm 40$, &$-32  \pm 2 $\\ 
               &&   &  &  &
                                                                  ($1320 \pm 20$) & \\

%        VIMOS & 28 June 2011  & $2070\pm 20$ & - & $1220 \pm 70$ & $40 \pm 1$ \\ 
 \hline \multicolumn{7}{|l|}{$\mathrm{H}\beta$ $\lambda4861.327$} \\
       &&&&& \\
 Xshooter & 1, 2, 3 March 2012 &17& $2260 \pm 30$ & $-30 \pm 5$ &$1370
                                                                  \pm
                                                                  10$
                                                            & $\sim -20 $\\ 
 \hline
    \end{tabular}
    \end{table*}
    
Using VIMOS spectroscopy,
\citet{2014MNRAS.440..365T} measured a significant radial velocity of
$ -170 \pm 20$ \kms\, in both the centroid of the $\mr{H}\alpha$ profile and photospheric absorption lines. This result is in contrast
with the low radial velocities obtained from the X-shooter data. To search
for any systematic effects that could alter the velocity
determinations, we  reanalysed the VIMOS spectra finding a radial
velocity consistent with that reported in \citet{2014MNRAS.440..365T}.  
The multi-slit mask of the VIMOS observations of CX1004 was designed to
have the counterpart centered on a $1\arcsec.0$ width slit. We
examined the acquisition images and the spatial profile for the
spectrum to confirm if that was the case. We found that the source 
was offset towards the North-East direction. This
positional offset from the center of the slit will have introduced a
significant radial velocity offset. Furthermore, light
from a field star $1\arcsec.3$ N-E from CX1004 fell inside the slit
contaminating its spectrum\footnote{A finding
chart is available in Appendix B in Torres et al. (2014).}. 
Since we cannot quantify the velocity offset with the available data,
we deem the radial velocities from the VIMOS spectra unreliable. 

A radial velocity of $-15\pm 20$ \kms was measured from GMOS
spectroscopy using a double-Gaussian fit to the averaged $\mr{H}\alpha$ profile (\citet{2015MNRAS.448.1900W}). To refine this determination and extract information from
individual profiles, we corrected the GMOS spectra for wavelength
zero-point offsets using the [OI] $\lambda6300.304$ sky emission line.
We recalculated the line parameters for the weighted average
data as done above for the X-shooter data. In this
  case, during the fits we applied a mask centered on the double--peaked emission line
with a width of the GMOS spectral resolution to exclude regions
affected by the unresolved line component from the M2 star.  The
results are shown in Table \ref{tab:tabFWHM}. The
  individual profiles were fit with a single Gausian model without
  masking the H$\alpha$ absorption line from the M2 star. The fits delivered a mean FWHM of 2430 \kms with r.m.s =
140 \kms, consistent with the value obtained from the same fitting
procedure applied to the averaged profile ($2510\pm40$ \kms). We also
cross--correlated the individual profiles against a Gaussian function with FWHM = 2450
\kms\, and calculated using a 2-Gaussian fit function the RV of the
H$\alpha$ centroid on the best quality individual data. None of the
methods delivered  evidence for significant variations in the line parameters.  

\subsection{Light curve analysis and astrometric matching of multi-band counterparts}
\label{sec:OGLElc}
\begin{figure}
\centering \includegraphics[width=1\columnwidth]{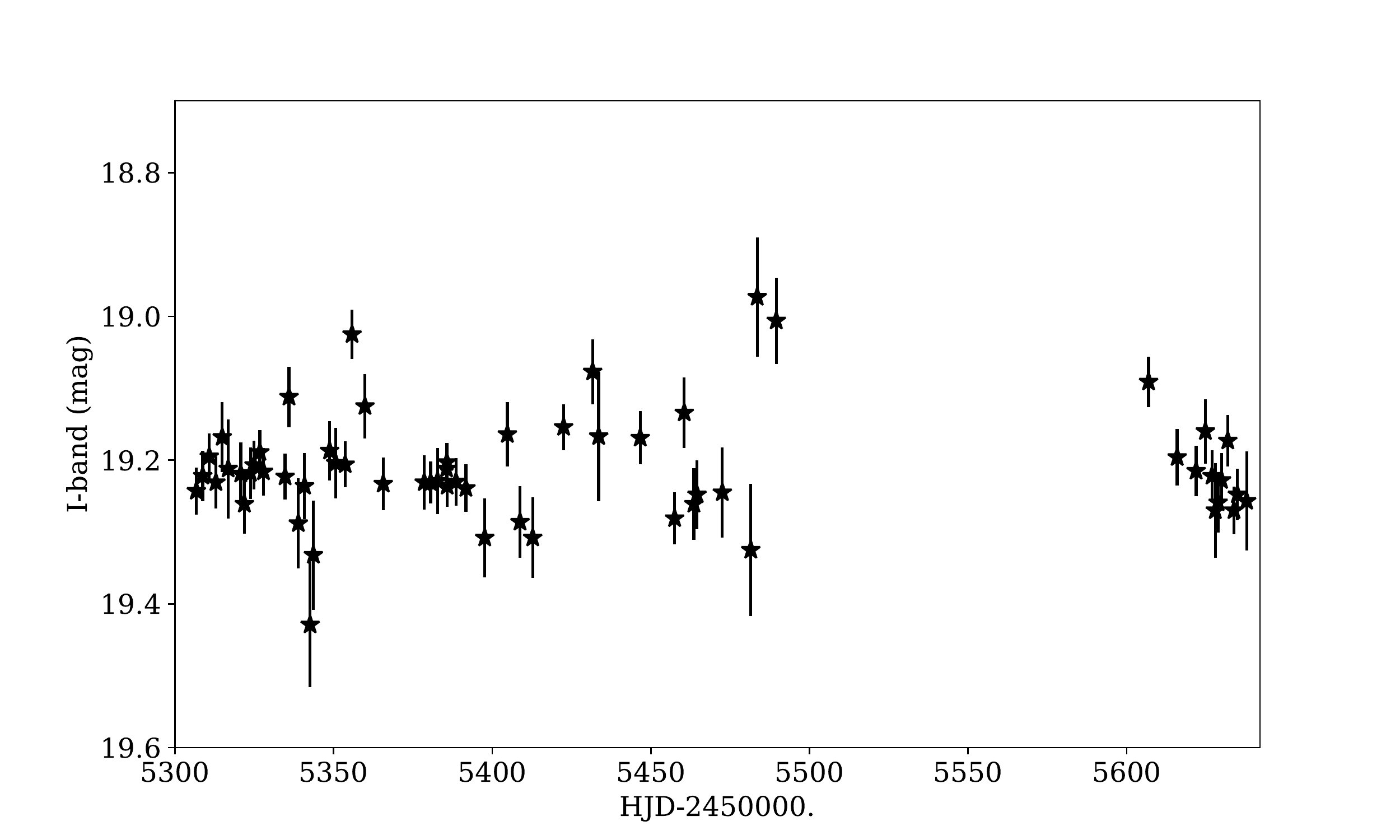}
\caption{23 April 2010 - 19 March 2011  OGLE I-band light curve of CX1004.}
\label{fig:ogle}
\end{figure}

In Fig. \ref{fig:ogle} we present the long-term (April 2010 - March
2011) I-band OGLE light curve for CX1004. The source has an average
magnitude of I$_\mr{OGLE}$ = $19.21$,  with an r.m.s. of 0.08 mag and (after rejecting photometric points with
SNR $< 15$) $\Delta \mr{I} = \mr{I}_\mr{max} - \mr{I}_\mr{min} = 0.3$. The light curve of field stars with similar I-band
brightness have comparable r.m.s. and
photometric errors ($\sim0.05$ mag on average). We therefore find no
evidence for photometric variability in the OGLE data. The field star $1\arcsec.3$ to the North-East from
CX1004 has an average magnitude I = $20.1$ and r.m.s. of 0.1 mag.
In our search for long-term variability we calibrate the  June 2015 FORS2 I-band PSF
photometry of both CX1004 and the N-E field star (section \ref{sec:fors2data})
using the I-band magnitudes of field stars in the
OGLE-IV optical source catalogue (\citealt{2015AcA....65....1U}). We
obtain I$_\mr{FORS2} = 19.21 \pm 0.02$ and $20.14 \pm 0.02$, for the
counterpart of CX1004 and the nearby field star, respectively, fully consistent with the OGLE photometry.

\begin{figure}
\centering \includegraphics[width=1\columnwidth]{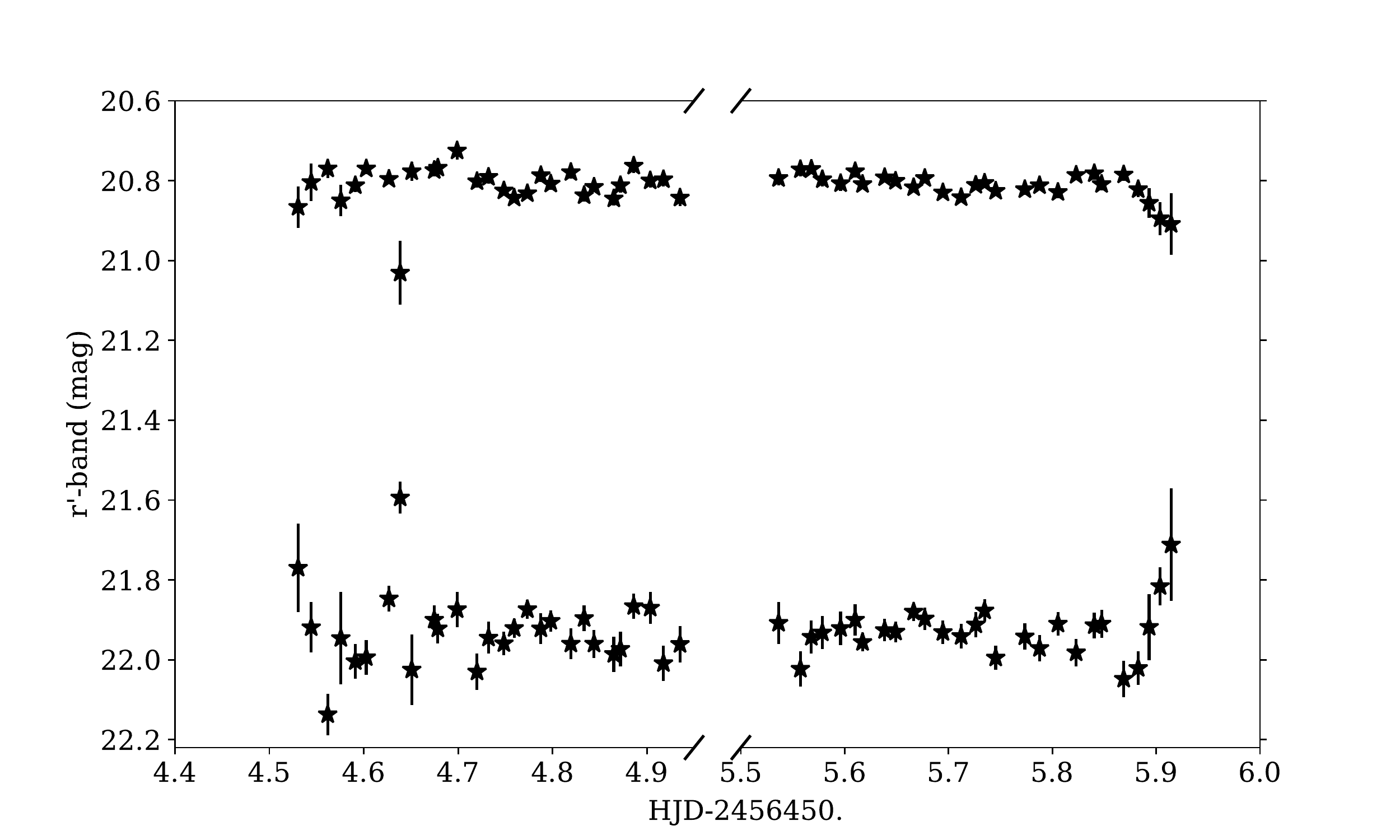}
\caption{DECam $r^\prime$-band light curve of CX1004 (top) and N-E
  field star (bottom). Note that the first, eight and final data points in both light
  curves are affected by correlated variability.}
\label{fig:DEC}
\end{figure}

We show in Fig. \ref{fig:DEC} the 10/11 June 2013 $r^\prime$-band DECam light curve of
CX1004 and the N-E field star. After removing the photometric points with
false (correlated) variability and following \citet{2016MNRAS.462L.106W},  we calculated periodograms
 over a period ranging from minutes to days. We find the light curve
of CX1004 to be aperiodic and we measure its  average $r^\prime$
magnitude, r.m.s. \,scatter, and maximum brightness variation
to be 20.80,
0.02 and 0.12 mag, respectively.  For the field star we obtain a mean magnitude of
21.94  and an r.ms.\, of  0.06 mag. In comparison, the GBS optical source catalogue (\citealt{2016MNRAS.458.4530W}) reports $r^\prime = 20.75 \pm  0.02$
from the 2006 observations of CX1004. When recalibrated with respect to the same
stars used for the DECam data and after rejecting points with
SNR $<15$, the 12-18 July 2010
Mosaic-II light curve \citep{2014ApJS..214...10B} has a mean  $r^\prime = 20.83$,
an r.m.s. of 0.04, and a $\Delta r^\prime = 0.16$. The small discrepancy
with respect to the 2006 Mosaic-II photometry is likely
 due to the fact that the N-E star was not resolved in the latter. Thus
 there is no evidence for intrinsic variability of the source.

The lack of radial velocity and photometric
variations implies that the M-type star is not associated with the
$\mr{H}\alpha$-emitting source. This can mean that this star is an
interloper unrelated to the X--ray binary or it might be the outer
star in a hierarchical triple system. 
Thus we looked for potential positional offsets  between multi-band optical
counterparts. For this search we analyzed the aligned B- and I-band
FORS2 images of the field (section \ref{sec:fors2data}). Aperture
photometry was performed for 103 point sources within a $\approx
500$\arcsec radius from the target to derive their emission centroid
in image coordinates. For each source, X and Y offsets were computed
and added in quadrature after converting them into arcsec. The offset
between the I- and B-band counterparts to CX1004 is $0\arcsec.05$. The
fraction of the 103 sources with a larger offset is $6\%$.  We also performed PSF
photometry on both images using {\small\texttt{DAOPHOT}}. We find no
evidence for underlying sources nor residuals in any of the images
after subtracting the PSF model. These results imply that in the
above photometric bands and at the time of the observations, the M-type star is
the dominant source of light making it impossible to test the
interloper scenario with our images.  \\

Finally, by
inspecting the archival VVV images,
we detect a uncatalogued point-like source that
matches the astrometric position of the optical counterpart to CX1004.
To compute its instrumental $K K_\mr{s}$--band magnitude and that of other point-like
objects in the field, we performed PSF photometry with
{\small\texttt{DAOPHOT}}. Differential
photometry with  respect to a nearby and non-variable field star
calibrated in VVV yields  $K_\mr{s} = 16.6 \pm 0.2$ for  CX1004. The
counterpart is not detected in the UKIRT Infrared Deep Sky  Survey
due to a poor quality of the data for CX1004. In addition, we note that the
$K_\mr{s} \approx 14.2$ candidate counterpart VVV
J174623.57-310550.75  reported in \citet{2014MNRAS.438.2839G}  is a nearby field star unrelated to CX1004. 

\subsection{Extinction and distance}
\label{sec:Reddening}
In this section we will employ the median values
1:1.85:13.44 for the ratios  A$_\mr{V}$:A$_\mr{I}$:A$_\mr{K_s}$
towards the Bulge \citep{2016MNRAS.456.2692N}, 
the ratio A$_\mr{K_s}$/E(J-K$_s$) = 0.528 \citep{2009ApJ...696.1407N} and the three--dimensional E(J-K$_s$) map
of the Galactic Bulge by \citet{2014A&A...566A.120S}. In addition,
 we will use the VISTA K${_s}$ $\simeq  \mr{K} + 0.044$ transformation (\citealt{2018MNRAS.474.5459G}, \citealt{2001AJ....121.2851C}) to convert between VISTA and Bessell \& Brett photometric systems.

We derive an estimate of the extinction A$_\mr{K_s}$ using the EW of the DIB at $\lambda8621$ which we measured in section \ref{sec:abs}.
The relation A$_\mr{K_s} \approx 0.691 \times
\mr{EW}(\mr{\AA})$ for this DIB 
(\citealt{2016MNRAS.463.2653D}) yields A$_\mr{K_s} =  0.13 \pm 0.04$ from
which we obtain A$_\mr{I}=1.0\pm 0.3$.
We can also estimate the reddening from the
observed (I-K$_s$) colour of the source assuming that the $90\%$
of the I-band brightness (i.e. I = 19.3) and the total infrared emission
(K$_s$ = $16.6 \pm 0.2$) is due to the light from the M2-type star
only. In this way we find an observed colour (I-K$_s$) $=2.7 \pm 0.2$,
redder than the intrinsic colour (I-K$_s$)$_0$ of 2.02 (2.22) for M2(M3) dwarfs (\citealt{1991AJ....101..662B}). 
Using A$_\mr{I}$/A$_\mr{K_s}=7.26$ and the colour excess, we estimate
A$_\mr{I}= 0.8 \pm 0.2$ consistent with the value calculated from the
DIB. Taking the I-band absolute magnitude for an M2/M3 V star ($\mr{M}_\mr{I}$
=8.00/8.71; \citealt{1991AJ....101..662B}) and adopting $\mr{A}_\mr{I}= 0.9$,
we find a distance of $\approx$ 1.8 / 1.3\kpc.  We can also
constrain the distance using the three--dimensional E(J-K$_s$) maps of the Bulge:
 the map towards CX1004 delivers E(J-K$_s$) = 0.21 and 0.31
 (i.e. A$_\mr{I}$ = 0.8 and 1.2) for a distance of 1.0 and 1.5 \kpc.
This is in line with the values found above for the distance which in turn
supports a main sequence luminosity class for the M2-type star.

\begin{figure}
\centering \includegraphics[width=1\columnwidth]{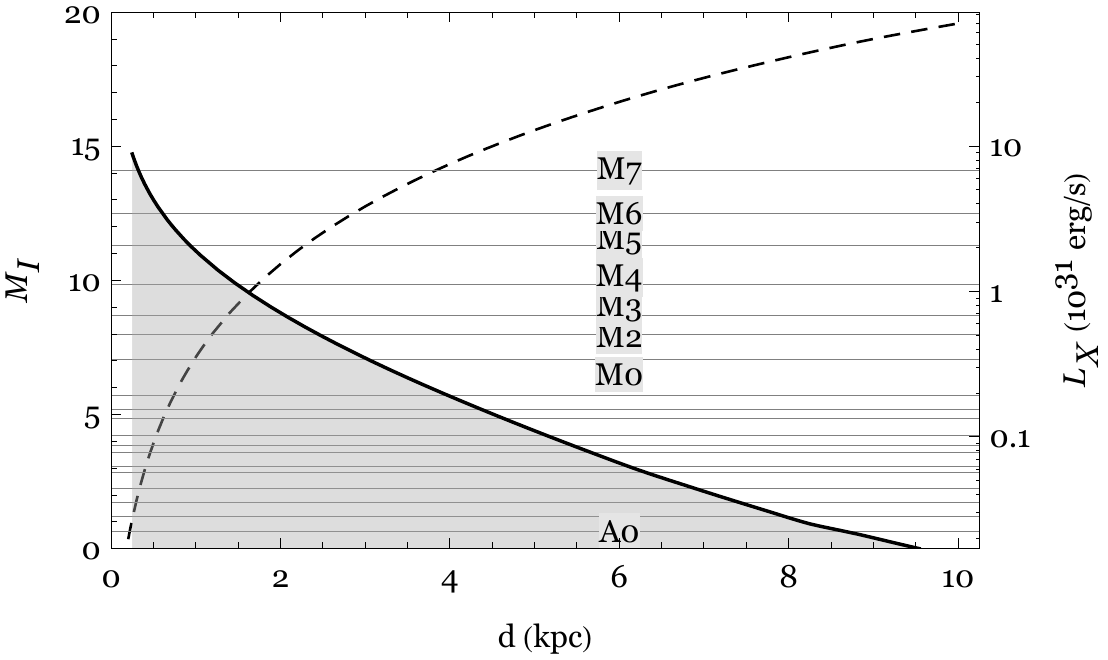}
\caption{Lower-limit for the I-band absolute magnitude of the putative
  donor star in CX1004 (solid line; left vertical scale) and expected X-ray luminosity
  as a function of distance (dashed line; right vertical scale). The area below the solid
  line is excluded as the object would be brighter in
  the I-band than observed. Grey horizontal lines mark the absolute
  magnitudes of M-type dwarfs and of K to A-type dwarfs in three subclasses: 0, 2 and 5.}
\label{fig:2axis}
\end{figure}

Given that the M2-type star is not the donor star in CX1004, we
derive I $\gtrsim 21.7$ for the accreting X-ray binary when
accounting for the fact that this contributes $\lesssim10\%$ to the 19.2 mag
I-band magnitude measured in section \ref{fig:Halpha}.  With this
constraint we can set  lower limits to the absolute I-band magnitude of the accreting
binary as a function of its distance. To do this, we first calculate
A$_\mr{I}$ as a function of the distance utilizing A$_\mr{I}$/A$_\mr{K_s}=7.26$,
$A{_\mr{K_s}}$/E(J-K$_s$) = 0.528 and  the
three--dimensional E(J-K$_s$) map in the direction
of CX1004. In Fig. \ref{fig:2axis} the solid line shows the resulting 
absolute I-band magnitude M$_\mr{I}$ of the X-ray binary as a function
of the distance. M$_\mr{I}$ also represents a lower limit to the absolute magnitude of the donor star since it does not 
account for the accretion disc contribution to the total light or
possible contamination from the field star $1\arcsec.3$ N-E of CX1004. The horizontal
lines in the figure mark the I-band absolute magnitude for M-dwarfs
(\citealt{1991AJ....101..662B}) and for A to K dwarfs (with subclasses
0, 2 and 5 for each spectral type; \citealt{2000asqu.book.....C}). We
also show with a dashed line in Fig. \ref{fig:2axis}  the X-ray
luminosity $L_X$  as a function of  distance. To estimate $L_X$ we
made use of {\small{\texttt{PIMMS}}} to convert the three $0.3-8$ keV
{\it Chandra}--detected X--ray counts of CX1004 to the unabsorbed $0.5-10$ keV  flux. 
In this calculation we assumed an absorbed power-law
spectrum with index $2.1$, which is typical for BH LMXBs in
quiescence (\citealt{2013ApJ...773...59P}) The
absorbing hydrogen column density $N_\mr{H}$ was obtained using the
empirical law $N_\mr{H}\approx 2.21 \times 10^{21} A_\mr{V}$ (\citealt{2009MNRAS.400.2050G}) and $A_\mr{V}/A_\mr{I}=1.85$. Note that this calculation of $L_x$ is subject to a
large margin for uncertainty given the  statistical noise in the
low-count detection of the source.

\section{Discussion}
\label{sec:discussion}

We have shown that the M2/3-type star that is astrometrically
consistent with the position of CX1004 and the H$\alpha$ emission line
source is at a distance of $1.3 - 1.8$ \kpc. If the actual X-ray
binary was at the same distance (whether in a hierarchical triple
system with this star or not), its donor will have to be of lower mass
than an M4 dwarf to contribute $\lesssim10 \%$ to the I-band
flux. Given this constraint on the spectral type of the donor, we can
estimate the orbital period (P$_\mr{orb}$) of the accreting binary
by employing the relation between the donor star mean
density ($\bar{\rho}$) and the orbital period for a Roche lobe filling star:
$\mr{P_\mr{orb} (hr)} \approx \sqrt{110/\bar{\rho} (\mr{g cm}^{-3})}$
(\citealt{2002apa..book.....F}). Using the empirical equations for the
mass and radius of low-mass stars by \citet{2016ApJ...819...87M} we
derive $\bar{\rho} \approx 21$ g cm$^{-3}$ for an M4 dwarf, thus
P$_\mr{orb}$ $\lesssim 2.3$ hr. In the case that the X-ray binary is
not located at the same distance as the M-dwarf, this cannot be much
further away from us since the reddening towards the source is such
that does not preclude the detection of the H$\gamma$ emission line in
the averaged spectrum. Given that we cannot quantify the exact value of the
  reddening we show here the magnitude of its effect on the inferred
  optical apparent magnitudes for two test distances. The three--dimensional reddening map towards CX1004 yields
  A$_\mr{V} \approx 3.7 (4.4)$ mag for a distance of 2.5 (3.0) \kpc. Taking the upper limit to the source brightness of
  I=21.7,  the above extinction already implies V=25.4 (26.1) for a
  conservatively adopted A0 spectrum. The reddening in the
  B-band ($\lambda_\mr{eff}$ = 4361 \AA), where the H$\gamma$ emission
  line is still distinguishable, will be even larger. If the X-ray
  binary is more distant, the combined effect of the larger reddening and distance make it quickly impossible to have detected the H$\gamma$ and H$\beta$ emission lines. The above evidence for  moderate reddening implies that the X-ray binary has to lie in front of the Galactic Bulge and supports a late-type dwarf mass donor
  (Fig. \ref{fig:2axis}). We also find it likely that the binary
is accreting at a low rate given the lack of evidence for He{\sc ii}
$\lambda4686$ emission in the spectra and the X-ray
luminosities estimated for distances $<4$ \kpc~
(Fig.\ref{fig:2axis}). These are  consistent with the 0.5 - 10 keV 
X-ray luminosities observed for both quiescent $\lesssim 4$ hr orbital
period  LMXBs ($\approx 10^{30-32}\mr{erg~s}^{-1}$; e.g. Fig. 3 in
\citealt{2014MNRAS.444..902A}) and non-magnetic CVs ($\approx
10^{29-32}~\mr{erg~s}^{-1}$; \citealt{2013MNRAS.430.1994R}). If CX1004
were a CV, the absence of evidence for dwarf
nova outbursts in our light curves implies that either we missed them
or the source has a lower duty cycle and lower X-ray luminosity
compared to non-magnetic CVs that show frequent outburst activity
\citep{2015MNRAS.448.3455B}.  In the latter case CX1004 would most
likely belong to the WZ Sge class of CVs which is characterized for
having orbital periods $\lesssim1.4$ hr and median outburst recurrence
times of  12 yr \citep{2015PASJ...67..108K}. \\

Another indirect way of characterizing the accreting binary in CX1004
is by analyzing the $\mr{H}\alpha$ emission line properties.  By
comparing its FWHM to those measured in a number of selected low
accretion rate CVs and LMXBs, \citet{2014MNRAS.440..365T} suggested
that the $\mr{H}\alpha$ line broadening is consistent with CX1004
being either an eclipsing CV or a high-inclination LMXB. Recently,
\citet{2018MNRAS.473.5195C} has presented a comprehensive study on the
$\mr{H}\alpha$ emission line properties for CVs and quiescent
LMXBs. This study shows that only eclipsing CVs with very short
($\lesssim 2.2$ hr) orbital periods can have H$\alpha$ emission line
FWHM $\geq 2200$ \kms.  As shown in section \ref{tab:tabFWHM}, the
average H$\alpha$ emission line FWHM in CX1004 exceeds this
limit. Thus the accreting binary would either be {\it{i)}} an
eclipsing short orbital period CV, {\it{ii)}} a high-inclination short
orbital period neutron star LMXB or {\it{iii)}} a BH LMXB observed at
moderate orbital inclination.  At this point, we cannot discriminate
between the CV and LMXB interpretation. For the CV case, we constrain
the donor star to white dwarf mass ratio ($q$) to be $<0.22$ for
P$_\mr{orb}< 2.2$ hr by using the empirical exponential relation
between $q$ and P$_\mr{orb}$ reported in \citet{2018MNRAS.473.5195C}.
For the BH case, limits on the donor star to BH mass ratio $q$ can be
set using the correlation established for quiescent BH LMXBs between
this quantity and the H$\alpha$ double-peak separation (DP) to FWHM
ratio: $\log q = -23.2 \log (\mr{DP/FWHM}) -6.88$
\citep{2016ApJ...822...99C}.  Following \citet{2016ApJ...822...99C},
we calculated the FWHM and DP by fitting the averaged
  X-shooter and GMOS H$\alpha$ profiles with single and 2--Gaussian
  models, with the latter having Gaussian components with identical
  width and height. The resulting FWHM and DP values are reported in
Table \ref{tab:tabFWHM}. The mass ratio was
evaluated through Monte Carlo randomization  where DP/FWHM was treated
as being normally distributed about its measured value with standard deviation equal to its uncertainty. From the X-shooter and
GMOS data we constrain $q$ to $0.09^{+0.03}_{-0.02}$ and
$0.08^{+0.06}_{-0.03}$, respectively. The quoted uncertainties
correspond to 68\% confidence level regions. For an LMXB hosting a 10
(5) \Msun BH, the lower limit for $q$ implies a 0.5 (0.25) \Msun donor
star. This result implies that the possibility of CX1004 being a
triple system containing a inner BH binary accreting from a donor of
spectral type later than M4 (0.26 M$_\odot$) is restricted since it
requires a $<5$ \Msun BH and these seem to be rare
\citep{2014SSRv..183..223C}.

Finally, we derive the radial velocity semi-amplitude of the donor
star (K$_\mr{d}$) using the H$\alpha$ FWHM - K$_\mr{d}$ correlation
for quiescent CVs and BH LMXBs \citep{2015ApJ...808...80C}. K$_\mr{d}$
is better evaluated when accounting for time variability of the line
profile. Therefore, we use for the FWHM the value of 2350 \kms and
r.m.s. = 140 \kms\, as measured from the GMOS data (section
\ref{sec:emission} ). We obtain K$_\mr{d}$ to be $=400 \pm 40$
\kms\,(CV scenario) and $=550 \pm 40$ \kms~(BH scenario).

\section{Conclusions}
\label{sec:conclusions}

The photometric data and the broad double-peaked
$\mr{H}\alpha$ emission line present in its spectrum allow CX1004 to
be a $\lesssim 2.2$ hr orbital period eclipsing CV or a
LMXB. However, neither radial velocity variations
nor line broadening is detected in the photospheric lines of the M2/3
dwarf optical counterpart. This implies that the M2/3-type star is not
the donor star of the accreting binary since radial variations should
have been measurable with our observations given the expected radial velocity semi-amplitude K$_\mr{d}$ ($\simeq 400 - 550$ \kms). The
available spectroscopic and photometric data do not allow us to
determine whether the accretor is a white dwarf, neutron star or a
black hole nor to discriminate between the two possible
interpretations for the M2/3 dwarf: an interloper along the line of
sight to the accreting binary or the outer companion in a triple
system. We note that two candidate hierarchical
  triple systems with red dwarf outer companions have been reported in
  the literature: the 3.5 hr orbital period CV RR Pic
  \citep{2017PASP..129a4201V} and the LMXB MAXI J1957+032
  \citep{2017ApJ...851..114R}. The best strategy to constrain the
compact object nature in CX1004 is to obtain the orbital period
through a radial velocity study of the H$\alpha$ emission line. For
this purpose higher SNR, higher time-resolution and continuous
coverage of the potential short orbital period are needed.  In
addition, radial velocities measured from the photospheric lines
will serve to further test the interloper scenario for the M2/3 star.

\section{Acknowledgments}
Based on observations made with ESO Telescopes at the La Silla Paranal
Observatory under programme ID 088.D-0096(A) and 095.D-0973(A). This
research has made use of the SIMBAD database, operated at CDS,
Strasbourg, France, and of the NASA's Astrophysics Data System. 
{\small{\texttt{IRAF}}} is distributed by the National Optical Astronomy Observatory, which is operated by the Association of Universities for Research in Astronomy (AURA) under a cooperative agreement with the National Science Foundation. 
Tom Marsh is thanked for developing and sharing his package
{\small{{\texttt{MOLLY}}}}. We thank the anonymous
  referee for constructive comments. MAPT and JC acknowledge support by the Spanish Ministry of
Economy, Industry and Competitiveness (MINECO) under
grant AYA2017-83216-P.
MAPT also acknowledges support via a Ram\'on y Cajal Fellowship (RYC-2015-17854).
The work of SR was supported by the Netherlands Research School for
Astronomy (NOVA).  PGJ and ZKR acknowledge support from
European Research Council Consolidator Grant 647208.
RIH and CBJ acknowledge support from NASA through Chandra Award Number AR3-14002X issued by the Chandra X-ray Observatory Center, which is operated by the Smithsonian Astrophysical Observatory for and on behalf of the National Aeronautics Space Administration under contract NAS8-03060.
The work of LW has been supported by the Polish National Science
Centre grant no. DEC-2011/03/B/ST9/02573. The OGLE project has received funding from the National Science Centre,
Poland, grant MAESTRO 2014/14/A/ST9/00121 and 2015/18/M/ST9/00544 to
Andrzej Udalski.  COH is supported by NSERC Discovery Grant
RGPIN-2016-04602, and a Discovery Accelerator Supplement. SR is thankful to Hagai Perets for his insights on the triple scenario.

\bibliographystyle{mn2e} \bibliography{mybib}{}

\label{lastpage}

\end{document}